\documentclass[11pt]{article}

\usepackage{amsmath,amsthm, amssymb,amsfonts}
\usepackage{epsfig,cancel}

\numberwithin{equation}{section}

\newcommand{\beq}{\begin{equation}}
\newcommand{\eeq}{\end{equation}}
\newcommand{\ba}{\begin{array}}
\newcommand{\ea}{\end{array}}
\newcommand{\bea}{\begin{eqnarray}}
\newcommand{\eea}{\end{eqnarray}}
\newcommand{\bean}{\begin{eqnarray*}}
\newcommand{\eean}{\end{eqnarray*}}

\newcommand{\comment}[1]{}

\newcommand{\cN}{{\cal N}}
\newcommand{\cF}{{\cal F}}
\newcommand{\cA}{{\cal A}}
\newcommand{\cB}{{\cal B}}
\newcommand{\cC}{{\cal C}}

\newcommand{\cV}{{\cal V}}

\def\cjn1{{\cA, \cC^*\otimes \wedge^j \cN^*}}
\def\bjn1{{\cA, \cB^*\otimes \wedge^j \cN^*}}
\def\vjn1{{\cA, \cV^*\otimes \wedge^j \cN^*}}
\def\cjn2{{\cA, \cC\otimes \wedge^j \cN^*}}
\def\bjn2{{\cA, \cB\otimes \wedge^j \cN^*}}
\def\vjn2{{\cA, \cV\otimes \wedge^j \cN^*}}

\def\fnote#1#2{\begingroup\def\thefootnote{#1}\footnote{#2}
     \addtocounter{footnote}{-1}\endgroup}

\begin{document}

\vspace{1cm}

\title{{\huge \bf The Edge Of Supersymmetry:\\
Stability Walls in Heterotic Theory }}

\vspace{2cm}

\author{
Lara B. Anderson${}^{1,2}$,
James Gray${}^{3}$,
Andre Lukas${}^{3}$,
Burt Ovrut${}^{1,2}$
}
\date{}
\maketitle
\begin{center} {\small ${}^1${\it School of Natural Sciences,
      Institute for Advanced Study, \\ Princeton, New Jersey 08540,
      U.S.A.} \\ ${}^2${\it Department of Physics, University of
      Pennsylvania, \\ Philadelphia, PA 19104-6395, U.S.A.}
    \\${}^3${\it Rudolf Peierls Centre for Theoretical Physics, Oxford
      University,\\
      $~~~~~$ 1 Keble Road, Oxford, OX1 3NP, U.K.}\\
    \fnote{}{andlara@physics.upenn.edu, j.gray1@physics.ox.ac.uk, lukas@physics.ox.ac.uk, ovrut@elcapitan.hep.upenn.edu} 
}
\end{center}

\abstract{We explicitly describe, in the language of four-dimensional
 ${\cal N}=1$ supersymmetric field theory, what happens when the
 moduli of a heterotic Calabi-Yau compactification change so as to
 make the internal non-Abelian gauge fields non-supersymmetric. At
 the edge of the region in K\"ahler moduli space where supersymmetry
 can be preserved, an additional anomalous $U(1)$ gauge symmetry
 appears in the four-dimensional theory.  The D-term contribution to
 the scalar potential associated to this $U(1)$ attempts to force the
 system back into a supersymmetric configuration and provides a
 consistent low-energy description of gauge bundle stability.}

\newpage

%
%

\section{Introduction}

Compactifications of the $E_8 \times E_8$ heterotic
string~\cite{Candelas:1985en,Green:1987mn} and heterotic
M-theory~\cite{Witten:1996mz}--\cite{Lukas:1998tt} on smooth, compact,
three-dimensional manifolds have been studied for many years.  The
simplest way to preserve ${\cal{N}}=1$ supersymmetry in the effective
four-dimensional theory is to choose the compactification space to be
a Calabi-Yau manifold, that is, to admit a metric with vanishing Ricci
tensor. In this paper, we will adopt this approach. These
compactifications necessarily involve the metric, $g_{a\bar{b}}$,
where $a,\bar{b}=1,2,3$ are the complex indices on the
threefold. Additionally, one must specify the gauge fields $A_{a}$,
with associated gauge group $G \subseteq E_{8}$, on the Calabi-Yau
manifold.  Whether or not these gauge fields preserve ${\cal{N}}=1$
supersymmetry in the effective theory is determined by studying the
variation of the higher-dimensional $E_{8}$ gauginos under
supersymmetry transformations. The gaugino variations vanish, and,
hence, ${\cal{N}}=1$ supersymmetry is preserved, if and only if the
gauge fields satisfy
\bea 
\label{HYM} 
F_{a \bar{b}} g^{a \bar{b}} = 0 \; , \; F_{ab} =
F_{\bar{a} \bar{b}}=0 
\eea  
where $F$ is the two-form field strength of the gauge field. These are known as the Hermitian Yang-Mills equations. 

Specifying a Calabi-Yau threefold, $X$, requires the choice of its
$h^{1,2}(X)$ complex structure moduli. These implicitly enter
\eqref{HYM} by defining the holomorphic and anti-holomorphic
coordinates. However, these moduli play no further role in this paper
and we will henceforth ignore them.  Crucially, we see that equations
\eqref{HYM} depend explicitly on the metric of the Calabi-Yau manifold
and, hence, on its $h^{1,1}(X)$ K\"ahler moduli. Indeed, whether these
equations even have a solution, that is, whether it is possible for
the gauge fields to preserve supersymmetry, is dependent on the values
taken by the K\"ahler moduli. Generically, K\"ahler moduli space
divides into regions where the gauge fields can preserve
supersymmetry and regions where they cannot~\cite{Sharpe:1998zu,Braun:2006ae,Braun:2006th}. At each point
in a supersymmetric region of K\"ahler moduli space the solution to Eqs.~\eqref{HYM}  depends on a number of arbitrary integration constants, the vector bundle moduli~\cite{Braun:2005fk}--\cite{Buchbinder:2002ji}. It is the combined K\"ahler/vector bundle moduli space in which we need to carry out our analysis.

For an Abelian internal gauge group, $G=U(1)$, it has been known for
some time~\cite{Distler:1987ee} that the supersymmetric part of the
K\"ahler moduli space is a locus of co-dimension one; that is,
effectively, one K\"ahler modulus is frozen if supersymmetry is to be
preserved. In the four-dimensional ${\cal{N}}=1$ effective theory,
this is described by a Fayet-Iliopoulos (FI)
D-term~\cite{Dine:1987xk}--\cite{Blumenhagen:2005ga} associated with a
$U(1)$ gauge symmetry which is anomalous in the Green-Schwarz
sense. In this paper, we will be concerned with {\em non-Abelian
  internal gauge groups}, specifically $G=SU(n)$. In this case, the
requirement that the internal gauge fields preserve supersymmetry does
not fix any of the K\"ahler moduli. However, it is known from a
mathematical analysis~\cite{Sharpe:1998zu,Braun:2006ae} using methods
of algebraic geometry, that the $h^{1,1}(X)$-dimensional K\"ahler
moduli space generically decomposes into subspaces; some in which the
non-Abelian internal gauge fields are supersymmetric, that is, where
they satisfy Eq.~\eqref{HYM}, and others where they break
supersymmetry.  Traditionally in the literature, the associated
four-dimensional effective theory is derived in the supersymmetric
region of moduli space where Eq.~\eqref{HYM} has a solution. In this
paper, we will extend this analysis and describe, for the first time
from a four-dimensional perspective, what happens as the moduli vary
and the gauge fields start to break supersymmetry.

Of central importance is the following observation: as the fields are
varied into the supersymmetry breaking region of moduli space, a new
potential will appear for the scalar fields of the four-dimensional
effective theory. As shown in Ref.~\cite{bigpaper}, a straightforward
dimensional reduction of the ten-dimensional Yang-Mills term and the
associated $R^2$ curvature term results in the following potential in
the four-dimensional theory,\footnote{This formula assumes that the field strength $F$ is
a $(1,1)$ form, even in the non-supersymmetric region where the first equation in Eq.~\eqref{HYM} cannot be solved.
In our context, the gauge fields are connections on holomorphic vector bundles for which this can always be arranged.}
\begin{eqnarray}
\label{finalintro}
V_{4d}=  \frac{\alpha'}{4}  \int_X \sqrt{-g} \left\{ {\rm Tr}(F^{(1)}_{a \bar{b}} g^{a \bar{b}})^2  + {\rm Tr}(F^{(2)}_{a \bar{b}} g^{a \bar{b}})^2 \right\}.
\end{eqnarray}
The notation here is standard \cite{Green:1987mn} with the field
strengths $F^{(1)}$ and $F^{(2)}$ being associated with the two $E_8$
factors in the gauge group and the integration taking over the
Calabi-Yau manifold.  For a supersymmetric field configuration satisfying
Eq.~\eqref{HYM}, the terms in the integrand of Eq.~\eqref{finalintro}
vanish. In this case, no potential is generated. However, if the
K\"ahler moduli are varied so that Eq.~\eqref{HYM} no longer has a
solution, that is, so that our gauge fields are no longer
supersymmetric, \eqref{finalintro} no longer vanishes and we obtain a
positive definite contribution to the potential energy as seen in 
four dimensions. We conclude that the region of moduli space in which the gauge connection satisfies \eqref{HYM} is {\it everywhere surrounded by a positive potential}. We refer to this as a {\it  ``wall of stability''.}

For the case of an Abelian internal gauge group, it can be explicitly seen that Eq.~\eqref{finalintro} leads to the FI D-term mentioned above. In the non-Abelian case, however, it might seem difficult to say more and present, for example, the potential as an explicit function of the four-dimensional moduli fields. After all, neither $F_{a \bar{b}}$ nor $g^{a \bar{b}}$ are known explicitly on a Calabi-Yau manifold. Nevertheless, an explicit form for this potential can indeed be derived. This is the main focus of this paper.

\section{Bundle Supersymmetry in K\"ahler Moduli Space: An Example}

It is clear from the preceding discussion that if an ${\cal N}=1$
supersymmetric subregion of the K\"ahler moduli space exists, it is
bounded by a positive definite potential wall, the boundary of
K\"ahler moduli space or both. It is known from a mathematical
analysis~\cite{Sharpe:1998zu,Braun:2006ae} that ``K\"ahler-cone''
sub-structure exhibiting both boundaries can exist. We now discuss
this in detail. For ease of exposition, we focus on a specific example
in the present paper, leaving the general construction to
Ref.~\cite{bigpaper}. The Calabi-Yau threefold, $X$, we consider is
defined as the zero locus of a polynomial of degree $(2,4)$ in the
homogeneous coordinates of an ambient space $\mathbb{P}^1 \times
\mathbb{P}^3$.  A common notation \cite{hubsch} for this manifold is
\begin{equation}
X= \left[\begin{array}[c]{c}
\mathbb{P}^1\\\mathbb{P}^3\end{array}\left|
\begin{array}[c]{ccc}2 \\4\end{array}\right.  \right]. \label{cicy}
\end{equation} 
This Calabi-Yau threefold has $h^{1,1}=2$ K\"ahler moduli and the K\"ahler form, $J$, can be written as 
\begin{equation}
J=t^1J_1+t^2J_2\; ,
\end{equation}
where $J_1$ and $J_2$ are the K\"ahler forms of $\mathbb{P}^1$ and $\mathbb{P}^3$ respectively, and $t^1$, $t^2$ denote the K\"ahler moduli. The K\"ahler cone, that is, the set of allowed K\"ahler moduli for this Calabi-Yau manifold, is characterised by $t^1 > 0$ and $t^2 > 0$. The K\"ahler moduli pair up with two axions, $\chi^1$, $\chi^2$, into the complex fields
\begin{equation}
T^{k}=t^{k}+2 i \chi^{k}, \qquad k=1,2\; .
\label{burt1}
\end{equation}
These form the bosonic parts of four-dimensional ${\cal{N}}=1$ chiral multiplets.
The axions descend from the M-theory three-form or, in the weakly coupled heterotic string, from the NS two-form.

\begin{figure}[!th]
  \centerline{\epsfxsize=2.5in\epsfbox{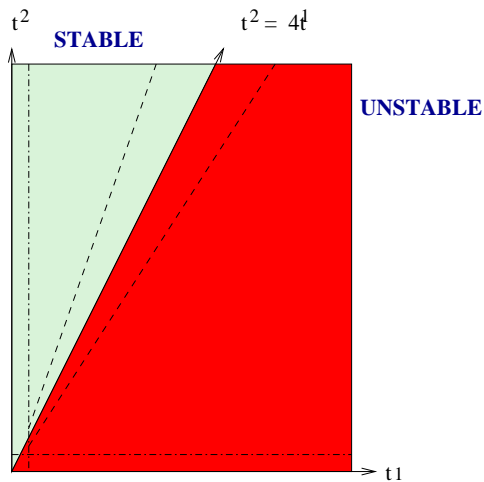}}
  \caption{\it\small The K\"ahler moduli space of the Calabi-Yau
    manifold~\eqref{cicy} in terms of the moduli
    $t^k=\textnormal{Re}(T^k)$. The allowed set of K\"ahler moduli (the
    K\"ahler cone) is the positive quadrant. The supersymmetric region
    where \eqref{HYM} admits a solution is marked in green (light
    shading), whereas the non-supersymmetric region where it does not
    is marked in red (dark shaded). The boundary between them is the
    line $t^{2}=4t^{1}$. The dash-dotted lines parallel to the axes
    indicate where supergravity breaks down as the K\"ahler moduli
    become too small. The additional $U(1)$ vector and Higgs
    supermultiplets are light compared to the compactification scale
    between the two dashed lines.}
\label{fig1}
\end{figure}

The gauge fields in this specific example are chosen as
follows. First, note that, mathematically, gauge fields are defined to
be connections on vector bundles which, in the present context, should
be holomorphic. We would like to construct such a holomorphic vector
bundle using line bundles on the Calabi-Yau manifold as building
blocks. For the above Calabi-Yau manifold, line bundles are
characterised by two integers, $k$, $l$, and are written as ${\cal
  O}_X(k,l)$. The first Chern class of these line bundles is given by
$c_1({\cal O}_X(k,l))=kJ_1+lJ_2$. The vector bundle we will consider
is a monad bundle, a construction which has been used in the physics
literature for some time~~\cite{Distler:1987ee},
\cite{Kachru:1995em}--\cite{Anderson:2008uw}. Such a bundle, $V$, is
defined as the kernel of a map, $f$, between two sums of line
bundles. The specific example we focus on in the present paper
is~\cite{Anderson:2007nc,Anderson:2008uw}
\bea 
\label{monad} 
0 \to V \to {\cal O}_X(1,0) \oplus {\cal O}_X(1,-1)
\oplus {\cal O}_X(0,1)^{\oplus 2} \stackrel{f}{\to} {\cal O}_X(2,1) \to 0\;. 
\eea 
The bundle $V$ has rank three and a vanishing first Chern class. Hence, it {\it generically} has a structure group $G=SU(3) \subset E_{8}$. What we would like to know is for which values of the K\"ahler moduli a gauge field connection on $V$ exists satisfying the Hermitian Yang-Mills equations \eqref{HYM} and, hence, preserving ${\cal N}=1$ supersymmetry? The general answer to this question was given by Donaldson, Uhlenbeck and Yau~\cite{duy} who proved the following theorem: {\it For a fixed choice of K\"ahler moduli, there exists a solution of the Hermitian Yang-Mills equations if and only if the vector bundle is ``slope-stable'', that is, has no destabilizing sub-bundles}.\footnote{The slope of a sub-bundle $\cF$ is defined as  $\mu(\cF)=\frac{1}{{\rm rank} \cF}\int_X c_1(\cF) \wedge J \wedge J$, where $J$ is the K\"ahler form. A bundle $V$ is slope-stable if and only if $\mu(\cF)<\mu(V)$ for all sub-bundles $\cF \subset V$. We also note that the slope can be explicitly expressed in terms of the K\"ahler moduli $t^i$ as $\mu ({\cF})=\frac{1}{{\rm rank}\cF} d_{ijk}c_1^i({\cF})t^jt^k$, where $d_{ijk}=\int_XJ_i\wedge J_j\wedge J_k$ are the triple intersection numbers of the Calabi-Yau manifold. For the Calabi-Yau manifold~\eqref{cicy}, the only non-vanishing intersection numbers are $d_{122}=4$ and $d_{222}=2$.}
We postpone a detailed analysis of slope-stability to~\cite{bigpaper} and simply state the result for the above bundle $V$. The ``maximally destabilizing'' sub-bundle for the bundle V in \eqref{monad} is a rank two bundle $\cF$ with first Chern class
\begin{equation}
 c_1(\cF )=-J_1+J_2\; . \label{c1F}
\end{equation}
It is easy to see that the slope of this sub-bundle is given by
\begin{equation}
 \mu (\cF )=(4 t^1t^2-(t^2)^2)\; . \label{muF}
\end{equation} 
This means that the slope is negative, and, hence, the bundle is supersymmetric, above the line $t^2=4t^1$, while supersymmetry is broken below this line.  In Fig.~\ref{fig1}, we have indicated these two regions in the K\"ahler cone of the Calabi-Yau manifold~\eqref{cicy}.
Finally, note that the exact sequence in \eqref{monad} actually defines a space of vector bundles, parametrized by a set of vector bundle moduli. In the supersymmetric region there are $h^{1}(X,V \otimes V^{*})=22$ such bundle moduli, which we denote by
\begin{equation}
\phi^{\alpha}, \qquad \alpha=1,\dots,22\; .
\label{burt2}
\end{equation}

In all previous heterotic literature, the four-dimensional
effective theory has been derived for K\"ahler moduli in the interior of the
green (light shaded) region of Fig.~\ref{fig1}, where the gauge fields preserve supersymmetry. In this paper, we will
extend the four-dimensional description so that we can explicitly
describe what happens as we approach and then cross over the $t^{2}=4t^{1}$ line, entering
the red (shaded) region where \eqref{HYM} cannot be solved.

\section{Four-Dimensional Effective Field Theory}

The first observation we make in deriving the four-dimensional field
theory is that, on the line between the supersymmetric and
non-supersymmetric regions in Fig.~\ref{fig1}, something special
must happen to the gauge fields. Although this is a line in K\"ahler
moduli space, if it is to support a solution to \eqref{HYM}, we find
that one is forced to a special locus in vector bundle moduli
space. As shown in Ref.~\cite{bigpaper}, the gauge fields must
``split'' into a direct sum; that is, whereas they were previously
valued in the adjoint of $SU(3)$, they now must become valued in the
adjoint representation of $S[U(2) \times U(1)]$ instead.  This latter
group is simply $U(2) \times U(1)$ where the determinant of the $U(2)$
matrix is constrained to be the inverse of the $U(1)$ phase.  The
gauge group of the four-dimensional theory is the commutant of the
structure group in $E_8$. When the structure group is $SU(3)$, which
it is in the supersymmetric part of the K\"ahler cone, this commutant
is $E_6$. However, when $SU(3)$ changes to $S[U(2) \times U(1)]$ on
the line between the supersymmetric and non-supersymmetric regions in
Fig. \ref{fig1}, the low energy gauge group is enhanced to $E_6
\times U(1)$. In summary: {\it On the line between the supersymmetric
  and non-supersymmetric regions, an additional $U(1)$ appears in the
  four-dimensional gauge group}.\footnote{In general, on a boundary
  wall in the K\"ahler cone, an $SU(n)$ bundle can decompose into
  $S[U(n_1)\times U(n_2) \times \ldots]$ where $\sum_i n_i =n$. In any
  such case, the commutant symmetry of $S[U(n_1)\times U(n_2) \times
  \ldots]$ in $E_8$ will always be enhanced by at least one $U(1)$
  symmetry. These, more general, cases are discussed in more detail in
  \cite{bigpaper}.} For the rest of this paper, we will work out the
four-dimensional effective theory including this additional $U(1)$. We
will show how this theory precisely reproduces the known physics in
the supersymmetric region of Fig. \ref{fig1}. We will then use it to
describe, for the first time, what happens in the non-supersymmetric
region, as well as the smooth transition between these two regimes.

A prerequisite to writing down the four-dimensional theory is a
knowledge of its spectrum. This can be calculated by computing the zero-modes of the Dirac operator twisted by the gauge fields. Equivalently, one can compute the cohomology of the various tensor products of the vector bundle $V$~\cite{Donagi:2004qk}--\cite{Donagi:2004ub}. At the stability wall, where the structure group of $V$ becomes $S[U(2) \times U(1)]$, the results are presented in Table \ref{table1}.
\begin{table}[h!]\begin{center}
\begin{tabular}{|c|c|c|}
 \hline
  \textnormal{Fields} & $E_6 \times U(1)$~\textnormal{charges} & \textnormal{number of fields} \\ \hline 
  $\psi^{\beta}$ & $1_{0}$ & 7 \\ \hline
 $B^I$ & $27_{-1/2}$ & 2 \\ \hline
  $C^L$ & $1_{-3/2}$ & 16 \\ \hline
\end{tabular}
\caption{\it\small The four-dimensional fields descending from 
higher-dimensional gauge fields. Shown are the vector bundle moduli,
  $\psi^{\beta}$, the $E_6$ ${\bf27}$-matter fields, $B^I$, and the
  $U(1)$ charged, $E_6$ singlets, $C^L$. The $U(1)$ charge of each
  field is shown as a subscript.}\label{table1}
\end{center}
\end{table}
A crucial observation is the following: {\it Only one sign of $U(1)$ charge appears in the low energy spectrum}. This will be of central importance in what follows. Naively, both signs of $U(1)$ charge could have appeared. However, it turns out that, in examples of the kind we are discussing here, the relevant Dirac operators never have $E_6$ singlet zero modes which are positively charged under $U(1)$. This will be explicitly proven in Ref.~\cite{bigpaper}.
The fields $B^I$ are the usual matter fields,
transforming in the ${\bf27}$ representation of $E_6$. There are no $\overline{{{\bf{27}}}}$ matter fields in the spectrum. Note that this model has only two generations. We have made no attempt to present a
phenomenologically viable example here. Rather, we have chosen our model to
be as simple as possible, while still illustrating the points we wish
to make. More complicated, phenomenologically realistic theories simply mimic the
structure we present here. The moduli $\psi^{\beta}$, $\beta=1,\dots7$ of the $S[U(2)\times U(1)]$ bundle can be thought of as a subset of the $SU(3)$ bundle moduli described earlier in \eqref{burt2}. Finally, we have the fields 
\begin{equation}
C^L, \qquad L=1,\dots,16. 
\label{burt4}
\end{equation}
These fields arise as a consequence of calculating the spectrum with
the gauge fields valued in the adjoint of $S[U(2) \times U(1)]$ rather
than $SU(3)$. Along with the K\"ahler moduli, they will play the key
role in the rest of our discussion. Internal gauge bundles with a
structure group that includes a $U(1)$ factor were constructed and
analysed for the first time in Ref.~\cite{Distler:1987ee}. It is also
known for some time~\cite{Lukas:1999nh,Blumenhagen:2005ga} that such a
$U(1)$ factor leads to an anomalous $U(1)$ gauge symmetry and an
associated $T$-modulus dependent FI D-term in the four-dimensional
$N=1$ effective theory. Here, we will apply these result to our theory
at the stability wall.

In addition to the fields of Table \ref{table1}, there are also the
usual heterotic moduli fields which include the complex structure and
K\"ahler moduli already mentioned, as well as the dilaton, $S$, and possible
M5 brane position moduli. How do these fields transform under the
additional $U(1)$? The complex structure moduli are invariant. The
imaginary parts of the K\"ahler moduli, dilaton and M5 brane position
moduli, in contrast, all transform under $U(1)$.  The $U(1)$
transformation of the K\"ahler moduli is of particular importance in
our context. It follows directly from the heterotic Bianchi identity
and is given by
\bea
\label{chiX}
\delta \chi^k = - \frac{3}{4} \; c^k_1(\cF ) \; \epsilon
\ , \eea
where the $T$-axions $\chi^{k}$ are defined in Eq.~\eqref{burt1}. Moreover, 
\begin{equation}
 c^{k}_1(\cF)=(-1,1) \label{c1comp}
\end{equation}
are the components of the first Chern class \eqref{c1F} with respect
to the basis $\{J_1, J_2\}$ of two-forms and $\epsilon$ is the
transformation parameter.  The result presented above is the lowest
order contribution to the transformation and will receive
one-loop corrections. These one-loop terms do not affect the present
discussion and will be neglected. However, their complete form is not
without interest and will be discussed in Ref.~\cite{bigpaper}.

It is known~\cite{Distler:1987ee,Lukas:1999nh} that low-energy $U(1)$ gauge
symmetries in heterotic compactifications which arise from the
presence of a $U(1)$ factor in the bundle structure group are
generally anomalous in a Green-Schwarz sense. In such cases, the
triangle anomaly is cancelled by the four-dimensional manifestation of
the Green-Schwarz mechanism. More precisely, this cancellation involves
a non-trivial transformation of the heterotic gauge kinetic
function under the $T$-axion
shifts~\eqref{chiX} \footnote{We emphasize that this is the case for a low-energy $U(1)$ gauge group that is also a factor in the vector bundle structure group~\cite{Dine:1987bq,Distler:1987ee,Lukas:1999nh}. For $U(1)$ symmetries for which this is not the case, the Green-Schwarz anomaly cancellation arises from the inhomogeneous shift of the dilatonic axion~\cite{Dine:1987xk}.}. For our example, it is immediately obvious that the $U(1)$ symmetry must indeed be anomalous
in this sense, since all of the $U(1)$ charges appearing in Table
\ref{table1} are negative.  As usual in the context of the
Green-Schwarz mechanism, the $U(1)$ vector supermultiplet picks up a
mass. From the point of view of the effective four-dimensional theory, this mass arises from a Higgs mechanism involving a linear combination of
all non-trivially transforming fields. We only consider situations where the $E_6$ part of the gauge group
is unbroken. Hence, $\langle B^I\rangle=0$ and the $B^I$ fields do not contribute to the $U(1)$ mass. However, the $T$ and the $C^L$ fields do contribute to the $U(1)$ mass, as we will later show.

We will now focus on the scalar potential of the four-dimensional
theory. Since we have ${\cal N}=1$ supersymmetry, this receives two
types of contributions; those from F-terms and those from D-terms. It
turns out that, in this simple example, only the D-term contributions
are important. The fact that all of the fields $C^L$ have the same
charge means that they cannot appear in a perturbative superpotential
in a gauge invariant manner. Since these are the fields which will be
important here, we will not consider F-terms further.  Using Table
\ref{table1}, Eq.~\eqref{chiX} and the standard formulas of ${\cal
  N}=1$ four-dimensional supergravity \cite{Wess:1992cp}, one can
write down the D-terms of our low energy theory. The $E_6$ D-terms are
completely standard in form.  Setting them to zero forces us to set
$\langle B^I \rangle =0$, consistent with the above assumption of an
unbroken $E_6$ gauge group. We will
thus discard these ${\bf 27}$ family fields in our subsequent
analysis. The $U(1)$ D-term is more interesting. We find, to quadratic order in the $C^{L}$ fields,  that
\bea 
\label{thedterm}
D^{U(1)} = f(t^i)+ \frac{3}{2} G_{L \bar{M}} C^L
{\bar{C}}^{\bar{M} } \ ,
\eea
where the FI term $f(t^i)$ is given by
\bea
\label{thedterm2}
f(t^i)=\frac{3}{4} \frac{\mu(\cF)}{\cal V} \; .
 \eea 
 Here, $G_{L \bar{M}}$ is the moduli space metric associated with the $C^{L}$ fields and is generically a function of the $t^{k}$ and $\psi^{\beta}$ moduli.  The only information we require about this quantity is that it is positive definite. 
${\cal V}$  is the Calabi-Yau volume which, in general,  is a cubic polynomial in the K\"ahler moduli $t^k$. For our specific example, it takes the form
\begin{equation}
 {\cal V}=2t^1(t^2)^2+\frac{1}{3}(t^2)^3\; . \label{V}
\end{equation} 
Additionally, $\mu(\cF)=4t^1t^2-(t^2)^2$ is the slope of the destabilising sub-bundle $\cF$, as in Eq.~\eqref{muF} . It is clear from the appearance of the slope in Eq.~\eqref{thedterm2} that the FI term is {\it positive in the non-supersymmetric (red, dark shaded)
region of Figure \ref{fig1}, negative in the supersymmetric (green, light shaded) region, and vanishes on the boundary line between these two}. The second term in (\ref{thedterm}) is the usual contribution to a $U(1)$ D-term from negatively charged fields and is positive semi-definite.

\begin{figure}[!th]
  \centerline{\epsfxsize=3in\epsfbox{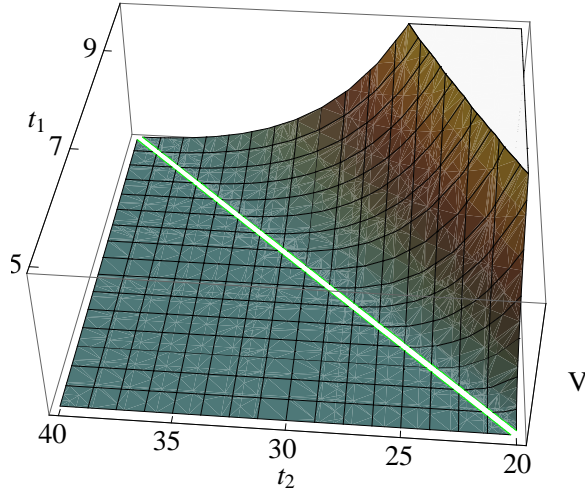}}
  \caption{\it\small The $U(1)$ D-term contribution to the scalar potential. The vertical axis is the potential V evaluated at the vacuum expectation values of the $C^{L}$ fields, while the horizontal plane is the K\"ahler moduli space shown in Figure \ref{fig1}. The line $t^{2}=4t^{1}$, where the slope $\mu(\cF)$ vanishes, indicates the stability wall which separates the supersymmetric and non-supersymmetric regions. Note that for ease of viewing, the axes have been rotated relative to Figure \ref{fig1}.}
\label{fig2}
\end{figure}

What happens to this D-term in the supposedly supersymmetric region of
K\"ahler moduli space? {\it For any K\"ahler moduli in the slope
  stable region, the FI term is negative and so can be cancelled by
  choosing suitable vacuum expectation values (vevs) for the fields
  $C^L$. Therefore, in this region, one can have $D^{U(1)}=0$. Hence, the vacuum energy  vanishes
  and supersymmetry is indeed preserved}. Note that there are sixteen fields,
$C^L$, but only one condition, $D^{U(1)}=0$, needed to specify the vacuum. 
Thus there are fifteen flat directions, that is, fifteen massless fields, in the
theory. {\it These fifteen massless fields plus the seven
  $\psi^{\beta}$ fields in Table \ref{table1} form the complete set of
  twenty-two vector bundle moduli $\phi^{\alpha}$ discussed in
  \eqref{burt2}. The one remaining field, which is massive and, generically, a vacuum dependent linear combination 
of the $T^{k}$ and $C^{L}$ fields, plays the role of the $U(1)$ Higgs chiral superfield.}

Let us examine which field gets a mass in more detail. Expanding all fields around a vacuum, 
that is, $t^{k}=\langle t^{k} \rangle +\delta t^{k}$ and $C^{L}=\langle C^{L} \rangle +\delta C^{L}$, with the vevs chosen so that
the D-term vanishes, we obtain the following expression to
first order in the field fluctuations and leading order in $\langle C^{L} \rangle$;
\bea \label{linearcombo} D^{U(1)} = -\frac{3}{4} G_{jk} c_1^j({\cal F}) \delta t^k + \frac{3}{2}
G_{L \bar{M}} \left( \left< C^L\right> \delta \bar{C}^{\bar{M}}
  + \delta C^L \left< \bar{C}^{\bar{M}} \right> \right) \ ,
\eea
where 
\begin{equation}
 G_{ij}=-\frac{\partial^2\ln{\cal V}}{\partial t^i\partial t^j}\;  \label{G}
\end{equation}  
is the K\"ahler moduli space metric, expressed in terms of the
Calabi-Yau volume ${\cal{V}}$ as given in (\ref{V}).  Note that here,
and henceforth, we will adopt the standard practice of denoting
$\langle t^{k} \rangle$ simply as $t^{k}$. The D-term contribution to
the potential is proportional to the square of this expression so that
the linear combination~\eqref{linearcombo} is, in fact, the Higgs
field. Its bosonic superpartner, the corresponding linear combination
of $T$-axions $\chi^i$ and phases of the fields $C^L$, is the
Goldstone mode which is absorbed by the $U(1)$ vector field. The Higgs
mass, $m_H$, which, from supersymmetry, must be equal to the $U(1)$
vector field mass $m_{U(1)}$ can be computed from
Eq.~\eqref{linearcombo} after canonically normalising the kinetic
terms $\frac{1}{4} G_{ij}\partial\delta t^i\partial \delta t^j$ and
$G_{L\bar{M}}\partial\delta C^L\partial\delta \bar{C}^{\bar{M}}$ of
the fields involved. Neglecting higher-order terms in $C$ and inverse
powers of the $T$-moduli one finds \bea
\label{u1mass} 
m^2_{H} =m_{U(1)}^2=\frac{1}{s}\left(\frac{9}{16} c_1^i(\cF)c_1^j(\cF)G_{ij} + \frac{9}{4} G_{L \bar{M}} 
\langle C \rangle^L \langle \bar{C} \rangle^{\bar{M}}\right) \; ,
\eea
where $s={\rm Re}(S)$ is the real part of dilaton.

To make this more concrete, let us determine the mass and corresponding Higgs multiplet at a point on the stability wall in K\"ahler moduli space. At the stability wall,  the slope of the
destabilizing sub-sheaf, ${\cal F}$, and, hence, the Fayet-Iliopoulos term, vanish. This means that the 
$C^{L}$ vevs also vanish in this
region of the vacuum space. Using this fact in Eq.~\eqref{linearcombo}, we
see that it is a particular combination of K\"ahler moduli, perpendicular
to the stability wall in K\"ahler moduli space, which becomes
massive on this locus. This can be explicitly verified for our example.
In this case, the field fluctuation~\eqref{linearcombo} of the D-term, evaluated
at the stability wall by setting $\langle C^L\rangle =0$ and $t^2=4t^1$, is
\begin{equation}
 D^{U(1)}=  \frac{9}{160}\frac{1}{(t^1)^2}(4\delta t^1-\delta t^2)\; . \label{Dex}
\end{equation} 
This shows that there is indeed one massive mode, given
by the linear combination $4\delta t^1-\delta t^2$, which represents the direction perpendicular to the stability
wall, as expected. The corresponding linear combination of axions $\chi^i$ is the Goldstone mode.
Evaluating Eq.~\eqref{u1mass} at the stability wall, one finds for the mass of the $U(1)$ vector field and the Higgs
\begin{equation}
m^2_{H} =m_{U(1)}^2= \frac{27}{128} \frac{1}{s(t^1)^2} \; . \label{u1masswall}
\end{equation} 
As one moves away from this line into the supersymmetric region, the Goldstone boson becomes a linear combination of K\"ahler axions and $C^{L}$ field phases. Further into the supersymmetric region, this becomes dominated by one of the $C^{L}$ phases. {\it We recover, therefore, the usual spectrum of heterotic theory, with two massless $T$-moduli, the usual vector bundle moduli and no additional scalars.  This is in agreement with the standard results for supersymmetric
heterotic compactifications.}

We would now like to argue that near the stability wall it is consistent to keep both the $U(1)$ vector supermultiplet and the Higgs chiral  superfield in the low-energy theory. For concreteness, we do this in the context of our specific example but the argument is of course general. Let us begin the analysis on the stability wall. The mass~\eqref{u1masswall} should be compared to the squared mass of a typical gauge sector massive mode which is of the order ${\cal V}^{-1/3}/s$. Using Eqs.~\eqref{V},  \eqref{u1masswall} leads to
\begin{equation}
 \frac{s\, m_{U(1)}^2}{{\cal{V}}^{-1/3}}\simeq \frac{1}{t^{1}} \;. \label{ratio}
\end{equation}
In the large radius limit, $t^1,t^2\gg 1$, we have $m^2_{U(1)}\ll {\cal{V}}^{-1/3}/s$ and the $U(1)$ mass is much smaller than typical heavy gauge sector masses on the stability line. Hence, in the regime where the supergravity approximation is valid it is always consistent to keep the $U(1)$ vector and Higgs supermultiplets in the low-energy theory. As a concrete example, for $t^{1} \gtrsim 10^2$, the vector and Higgs supermultiplets masses are at least an order of magnitude below the mass scale of heavy gauge states. What happens as we move away from the stability line into the supersymmetric region of moduli space? In this case, the vevs of the $C^{L}$ fields are no longer zero and the $U(1)$ mass, now given by expression (\ref{u1mass}), increases. Eventually it becomes of the order of the compactification scale and the vector and Higgs supermultiplets should no longer be kept in the low-energy theory. The region around the stability line for which $s\, m_{U(1)}^2/{\cal V}^{-1/3}<10^{-1}$  is indicated in Fig.~\ref{fig1}.

\vspace{0.1cm}

We can now go further and describe what happens in the
non-supersymmetric region of K\"ahler moduli space.  {\it In this
  region, the FI term given in \eqref{thedterm} is positive. Since the
  second term in \eqref{thedterm} is also positive, it is no longer
  possible to adjust the vevs of the $C^L$ fields to cancel the FI
  term. Therefore, we find that $D^{U(1)} \neq 0$ at every point in
  this region of moduli space, reproducing the fact that supersymmetry
  is broken in the four-dimensional effective theory. The
  square of the D-term in the red (dark shaded) region of Fig.
  \ref{fig1} gives rise to an everywhere positive four-dimensional
  potential}.  Note that this potential is minimized at each point
in K\"ahler moduli space by setting $\left< C^L \right>=0$. The
resulting potential for the K\"ahler moduli is plotted in
Fig. \ref{fig2}. Close to the stability wall in Fig. \ref{fig1}, this
potential is still relatively small and it makes sense to talk about a
four-dimensional theory. However, {\it sufficiently far into the
  unstable region the energy density of the potential becomes
  comparable to the compactification scale and one would expect that
  no four-dimensional description exists}. Since, in the absence of
other effects, there is no perturbative vacuum in the
non-supersymmetric part of the K\"ahler cone, we will refrain from
discussing masses in this region, except to note that the $U(1)$
vector supermultiplet continues to have the mass \eqref{u1mass}
induced by the Green-Schwarz mechanism.

\vspace{0.1cm}

To summarize: We have introduced a new D-term contribution to the
potential of heterotic string and M-theory which is positive
semi-definite and describes the supersymmetry properties of the
internal non-Abelian gauge fields from the perspective of the four-dimensional
effective theory.  This D-term originates from an anomalous $U(1)$
gauge symmetry which arises in the low-energy theory. The
corresponding vector supermultiplet has a mass induced by the
Green-Schwarz mechanism. Associated with this enhanced $U(1)$ are
charged light states in the spectrum. In the part of the K\"ahler
moduli space where the internal vector bundle is supersymmetric, these
states develop vacuum expectation values which cancel the
Fayet-Iliopoulos term.  In the part of K\"ahler moduli space where the
bundle breaks supersymmetry, the Fayet-Iliopoulos term changes sign
and can no longer be cancelled by vacuum expectation values of the
charged states. Thus, the D-term vanishes in the region of K\"ahler
moduli space where the internal gauge fields are supersymmetric, and
is non-vanishing where the internal gauge fields break
supersymmetry. For this mechanism to work, it is crucial that all of
the charged states have the same sign of $U(1)$ charge. We have
checked that this is indeed the case. Our picture provides, for the
first time, a concrete four-dimensional description of supersymmetry
breaking induced by non-Abelian heterotic gauge bundles.

The new potential we have described has many possible applications,
from cosmology to moduli stabilization. One might imagine using
non-perturbative effects to stabilise moduli a small way into the
non-supersymmetric region, thus obtaining a naturally small scale of
supersymmetry breaking. The global remnant of the $U(1)$ symmetry
described here constrains the Lagrangian in the supersymmetric
region. This may allow us to place restrictions on which
vector bundles lead to realistic particle phenomenology. There are
also more formal applications of our work. These  concern, for example, linking what
might be seen as different vector bundles in physical moduli space and
proving bundle stability purely  from four-dimensional field theoretical
arguments. The authors hope to explore such topics in future
publications.

\section*{Acknowledgments}
The authors would like to thank J.~Maldacena and N.~Seiberg for helpful discussions.
J.~G.~would like to thank the Department of Physics and Astronomy, University of Pennsylvania for hospitality
while some of this work has been completed. The work of L.A. and B.A.O. is supported in part by the DOE under contract No. DE-AC02-76-ER-03071.
A.~L.~is supported by the EC 6th Framework Programme
MRTN-CT-2004-503369.  J.~G.~is supported by STFC.


\end{document}